# A model for exchange-biased asymmetric giant magneto-impedance in amorphous wires

N.A. Buznikov[1,*], S.S. Yoon[2], C.G. Kim[1] and C.O. Kim[1]

[1] Research Center for Advanced Magnetic Materials, Chungnam National University, Daejeon 305-764, Republic of Korea

[2] Department of Physics, Andong National University, Andong 760-749, Republic of Korea

**Abstract**

A model describing the exchange-biased asymmetric giant magneto-impedance in Joule-heated amorphous wires is proposed. It is assumed that the Joule heating results in the formation of a surface hard magnetic crystalline layer, and the asymmetric giant magneto-impedance is related to the exchange coupling between the amorphous and crystalline phases. The coupling between the surface layer and the amorphous bulk is described in terms of an effective bias field. The model is based on a simultaneous solution of linearizied Maxwell equations and the Landau–Lifshitz equation. The calculated field and frequency dependences of the wire impedance are in a qualitative agreement with results of the experimental study of the asymmetric giant magneto-impedance in Joule-heated Co-based amorphous wires.

---

[*] Corresponding author. *E-mail address:* n_buznikov@mail.ru

# 1. Introduction

The giant magneto-impedance (GMI) is one of the most promising magnetotransport effects due to its application in highly sensitive magnetic-field sensors. The GMI effect consists of a huge change in the impedance of a magnetic conductor in the presence of a static magnetic field and has been observed in different soft magnetic materials, in particular, in amorphous wires, ribbons, glass-coated microwires and film structures [1]. The linearity and the sensitivity for the magnetic field are the most important parameters in practical applications of the GMI effect. To improve the linear characteristic of the GMI response, the asymmetric GMI is very promising [1,2].

There are several mechanisms of the asymmetric GMI [1]; one of these is related to the exchange coupling between two magnetic phases. This type of the asymmetric GMI has been observed in Co-based amorphous ribbons annealed in air in the presence of a weak magnetic field applied along the sample axis [3,4]. Since the ribbons annealed in vacuum did not show the asymmetry [3,5], the phenomenon has been attributed to oxidation and surface crystallization. The field-annealing in air produces asymmetric hysteresis loops in amorphous ribbons due to the exchange interaction of the amorphous bulk with the magnetically harder crystalline surface layers [6]. In the presence of the annealing field, unidirectional magnetic anisotropy is induced in the crystalline layers. Because the crystalline phase is the hard magnetic one, it remains magnetically ordered within a relatively wide range of magnetic fields. The exchange coupling between the crystalline and amorphous phases produces an effective bias field resulting in asymmetry in the dc magnetization of the amorphous bulk that is responsible for the asymmetric GMI in field-annealed ribbons [3,5,7,8].

Recently, the exchange-biased asymmetric GMI has been observed in $Co_{68.8}Fe_{4.32}Si_{12.5}B_{15}$ amorphous wires of diameter 115 μm subjected to Joule heating in open air [9]. Although the structure analysis of the wires has not been carried out, the magneto-static and magneto-impedance data confirm that Joule heating produces crystallization at the wire surface. The magnetic field of the annealing current induces circular unidirectional anisotropy in the crystalline layer, and the asymmetric GMI can be ascribed to the exchange coupling between the surface layer and the amorphous region. It should be noted that the configuration of the unidirectional anisotropy differs from that in experiments with field-annealed ribbons [3–5].

The aim of this paper is to present a model for analysis of the exchange-biased asymmetric GMI in Co-based amorphous wires. It is assumed that the wire consists of an amorphous region and a surface crystalline layer. The coupling between the crystalline layer and the amorphous bulk is described through an effective bias field. The solution of Maxwell equations together with the Landau–Lifshitz equation is found for the amorphous and crystalline regions neglecting a domain structure. The field and frequency dependences of the asymmetric GMI response are analysed. The results obtained allow one to explain the main features of the asymmetric GMI in Joule-heated Co-based amorphous wires observed in experiment [9].

## 2. Model

Let us consider a wire of diameter $D$ consisting of an amorphous bulk and a surface crystalline layer. The ac current $I=I_0\exp(-i\omega t)$ flows along the wire (along $z$-axis), and the external dc magnetic field $H_e$ is parallel to the current. The Joule heating results in the crystallization of the surface layer, and the magnetic field generated by the annealing current induces circular unidirectional anisotropy in the crystalline layer of thickness $t_c$. Since it is assumed that the crystalline phase is the hard magnetic one, the magnetization of the surface layer does not vary with the external magnetic field, and the magnetization direction coincides with that of the unidirectional anisotropy field $H_u$.

We neglect a domain structure in the wire and assume that the amorphous bulk has the helical uniaxial anisotropy, and the anisotropy axis makes the angle $\psi$ with the circular direction. The exchange coupling between the amorphous and crystalline phases induces the effective bias field $H_b$ in the amorphous region, with the bias field being in the opposite direction to the unidirectional anisotropy field $H_u$ [7,8]. For the analysis, we assume that the effective bias field $H_b$ does not vary over the amorphous region. Although this approximation simplifies significantly the real spatial distribution of the bias field, it allows one to find an analytical solution for the field distribution within the wire and to calculate the impedance.

Furthermore, to simplify calculations, it is assumed that the conductivity $\sigma$ and the saturation magnetization $M$ are the same for the amorphous and crystalline phases.

Under the assumption of a local relationship between the magnetic field and the magnetization and in a linear approximation with time-variable parameters, the distribution of the electric and magnetic fields within the wire satisfies Maxwell equations

$$\text{curl}\,\mathbf{e} = (i\omega/c)\hat{\mu}_k \mathbf{h}, \qquad (1)$$
$$\text{curl}\,\mathbf{h} = 4\pi\sigma \mathbf{e}/c,$$

where $\mathbf{e}$ and $\mathbf{h}$ are the amplitudes of the ac electric and magnetic fields, $c$ is the velocity of light and $\hat{\mu}_k$ ($k=1,2$) are the permeability tensors for the amorphous region and the surface crystalline layer, respectively. In the present analysis, we neglect a domain structure, and then the permeability tensor $\hat{\mu}_1$ in the amorphous bulk is determined only by the magnetization rotation. The tensor $\hat{\mu}_1$ can be found from a solution of the linearized Landau–Lifshitz equation by using the conventional methods of the ferromagnetic resonance theory. The permeability is represented by a tensor having six different parameters [10].

Taking into account that the fields depend only on the radial coordinate $\rho$, within the amorphous region, $\rho < D/2 - t_c$, the Maxwell equations can be reduced to two coupled differential equations for the circular, $h_\varphi^{(1)}$, and longitudinal, $h_z^{(1)}$, components of the magnetic field [10,11]:

$$\frac{\partial^2 h_\varphi^{(1)}}{\partial \rho^2} + \frac{1}{\rho}\frac{\partial h_\varphi^{(1)}}{\partial \rho} - \frac{h_\varphi^{(1)}}{\rho^2} + \frac{2i}{\delta^2}\times(1+\mu_1 \sin^2\theta)h_\varphi^{(1)} = \frac{2i}{\delta^2}\times \mu_1 h_z^{(1)} \sin\theta\cos\theta,$$
$$\frac{\partial^2 h_z^{(1)}}{\partial \rho^2} + \frac{1}{\rho}\frac{\partial h_z^{(1)}}{\partial \rho} + \frac{2i}{\delta^2}\times(1+\mu_1 \cos^2\theta)h_z^{(1)} = \frac{2i}{\delta^2}\times \mu_1 h_\varphi^{(1)} \sin\theta\cos\theta, \qquad (2)$$

where $\delta = c/(2\pi\sigma\omega)^{1/2}$ is the skin depth in non-magnetic material, $\theta$ is the equilibrium angle between the magnetization vector and the circular direction and $\mu_1$ is the effective permeability of the amorphous region, which is given by [10]

$$\mu_1 = \frac{\gamma 4\pi M}{\omega_1 - i\kappa\omega - \omega^2/(\omega_2 - i\kappa\omega)},$$
$$\omega_1 = \gamma[H_a \cos\{2(\theta-\psi)\} - H_b \cos\theta + H_e \sin\theta], \qquad (3)$$
$$\omega_2 = \gamma[4\pi M + H_a \cos^2(\theta-\psi) - H_b \cos\theta + H_e \sin\theta].$$

Here $H_a$ is the uniaxial anisotropy field in the amorphous region, $\gamma$ is the gyromagnetic constant and $\kappa$ is the Gilbert damping parameter. The equilibrium magnetization angle $\theta$ can be found by minimizing the free energy, which can be presented as a sum of the anisotropy energy, bias field energy and Zeeman energy. The minimization procedure results in the following equation for $\theta$:

$$H_a \sin(\theta - \psi)\cos(\theta - \psi) - H_b \sin\theta - H_e \cos\theta = 0. \tag{4}$$

The electric field components in the amorphous region can be found by means of equations

$$e_z^{(1)} = \frac{c}{4\pi\sigma}\left[\frac{\partial h_\varphi^{(1)}}{\partial \rho} + \frac{h_\varphi^{(1)}}{\rho}\right], \quad e_\varphi^{(1)} = -\frac{c}{4\pi\sigma}\frac{\partial h_z^{(1)}}{\partial \rho}. \tag{5}$$

Let us restrict our consideration to the case of high enough frequency, when the effective skin depth in the amorphous region is much less than the wire radius. In this case, the solution of equations (2) and (5) can be presented in the following form [11,12]:

$$\begin{aligned}
h_\varphi^{(1)}(\rho) &= A_1 \cos\theta \exp\{\lambda_1(\rho - d/2)\} + A_2 \sin\theta \exp\{\lambda_2(\rho - d/2)\}, \\
h_z^{(1)}(\rho) &= A_1 \sin\theta \exp\{\lambda_1(\rho - d/2)\} - A_2 \cos\theta \exp\{\lambda_2(\rho - d/2)\}], \\
e_z^{(1)}(\rho) &= (c/4\pi\sigma)[A_1\lambda_1 \cos\theta \exp\{\lambda_1(\rho - d/2)\} + A_2\lambda_2 \sin\theta \exp\{\lambda_2(\rho - d/2)\}], \\
e_\varphi^{(1)}(\rho) &= -(c/4\pi\sigma)[A_1\lambda_1 \sin\theta \exp\{\lambda_1(\rho - d/2)\} - A_2\lambda_2 \cos\theta \exp\{\lambda_2(\rho - d/2)\}].
\end{aligned} \tag{6}$$

Here $A_1$ and $A_2$ are the constants, $d = D - 2t_c$ is the amorphous region diameter, $\lambda_1 = (1-i)/\delta$ and $\lambda_2 = (1-i)(\mu_1 + 1)^{1/2}/\delta$. It follows from equations (6) that the high-frequency approximation is valid if $\delta/(\mu_1 + 1)^{1/2} \ll d/2$. Note that in the case of an arbitrary frequency, the solution of equations (2) can be found in the form of series representing even and odd types of solutions with respect to the radial coordinate [11].

Since the unidirectional anisotropy field $H_u$ in the crystalline layer is sufficiently high, the permeability in the surface layer is almost independent of the external field. The permeability tensor $\hat{\mu}_2$ in the crystalline layer can be written as

$$\hat{\mu}_2 = \begin{pmatrix} 1 + \mu_a & 0 & i\mu_b \\ 0 & 1 & 0 \\ -i\mu_b & 0 & 1 + \mu_a \end{pmatrix},$$

$$\mu_a = \frac{\gamma 4\pi M(\gamma H_u - i\kappa\omega)}{(\gamma H_u - i\kappa\omega)^2 - \omega^2}, \tag{7}$$

$$\mu_b = \frac{\gamma 4\pi M \omega}{(\gamma H_u - i\kappa\omega)^2 - \omega^2}.$$

Taking into account equation (7), the general solution of Maxwell equations (1) in the surface crystalline layer, $d/2 < \rho < D/2$, can be expressed as

$$h_\varphi^{(2)}(\rho) = B_1 J_1(k_1\rho) + B_2 Y_1(k_1\rho),$$
$$h_z^{(2)}(\rho) = C_1 J_0(k_2\rho) + C_2 Y_0(k_2\rho),$$
$$e_z^{(2)}(\rho) = (ck_1/4\pi\sigma)[B_1 J_0(k_1\rho) + B_2 Y_0(k_1\rho)],$$
$$e_\varphi^{(2)}(\rho) = (ck_2/4\pi\sigma)[C_1 J_1(k_2\rho) + C_2 Y_1(k_2\rho)]. \qquad (8)$$

Here $B_1$, $B_2$, $C_1$ and $C_2$ are the constants, $J_j$ and $Y_j$ ($j=0,1$) are the Bessel functions of the first and the second kind, respectively, $k_1 = (1+i)/\delta$, $k_2 = (1+i)\mu_2^{1/2}/\delta$ and the effective longitudinal permeability $\mu_2$ in the crystalline layer is given by

$$\mu_2 = \frac{(1+\mu_a)^2 - \mu_b^2}{1+\mu_a} = \frac{(\gamma 4\pi M + \gamma H_u - i\kappa\omega)^2 - \omega^2}{(\gamma 4\pi M + \gamma H_u - i\kappa\omega)(\gamma H_u - i\kappa\omega) - \omega^2}. \qquad (9)$$

The components of the electric and magnetic fields should satisfy the continuity conditions at the interface between the amorphous region and the crystalline layer. In addition, the components of the magnetic field at the wire surface are determined by the excitation conditions [10,11]. Thus, we have the following equations for the fields:

$$e_z^{(1)}(d/2) = e_z^{(2)}(d/2),$$
$$h_\varphi^{(1)}(d/2) = h_\varphi^{(2)}(d/2),$$
$$e_\varphi^{(1)}(d/2) = e_\varphi^{(2)}(d/2),$$
$$h_z^{(1)}(d/2) = h_z^{(2)}(d/2), \qquad (10)$$
$$h_\varphi^{(2)}(D/2) = 4I_0/cD,$$
$$h_z^{(2)}(D/2) = 0.$$

The six constants in equations (6) and (8) can be found from the set of equations (10). After that, the wire impedance $Z$ can be calculated by means of the usual relation [10–12]

$$Z = \frac{l e_z^{(2)}(D/2)}{I_0} = \frac{4l}{cD} \times \frac{e_z^{(2)}(D/2)}{h_\varphi^{(2)}(D/2)}$$
$$= \frac{k_1 l}{\pi\sigma D} \times \frac{B_1 J_0(k_1 D/2) + B_2 Y_0(k_1 D/2)}{B_1 J_1(k_1 D/2) + B_2 Y_1(k_1 D/2)}, \qquad (11)$$

where $l$ is the wire length. It should be noted that although the parameter $k_1$ is independent of the external magnetic field, equation (11) describes the field dependence of the impedance $Z$ through the constants $B_1$ and $B_2$.

Using the solution of equations (10), we can present expression for the impedance in the following form:

$$Z/R_{dc} = (k_1 D/4)$$
$$\times [F_1 k_1 \{k_2 G_1 + (\lambda_1 \sin^2\theta + \lambda_2 \cos^2\theta)G_2\} + F_2\{k_2(\lambda_1 \cos^2\theta + \lambda_2 \sin^2\theta)G_1 + \lambda_1\lambda_2 G_2\}] \quad (12)$$
$$\times [F_3 k_1 \{k_2 G_1 + (\lambda_1 \sin^2\theta + \lambda_2 \cos^2\theta)G_2\} + F_4\{k_2(\lambda_1 \cos^2\theta + \lambda_2 \sin^2\theta)G_1 + \lambda_1\lambda_2 G_2\}]^{-1},$$

where $R_{dc} = 4l/\pi\sigma D^2$ is the dc wire resistance and

$$\begin{aligned}
F_1 &= J_0(k_1 d/2)Y_0(k_1 D/2) - J_0(k_1 D/2)Y_0(k_1 d/2), \\
F_2 &= J_0(k_1 D/2)Y_1(k_1 d/2) - J_1(k_1 d/2)Y_0(k_1 D/2), \\
F_3 &= J_0(k_1 d/2)Y_1(k_1 D/2) - J_1(k_1 D/2)Y_0(k_1 d/2), \\
F_4 &= J_1(k_1 D/2)Y_1(k_1 d/2) - J_1(k_1 d/2)Y_1(k_1 D/2), \\
G_1 &= J_1(k_2 d/2)Y_0(k_2 D/2) - J_0(k_2 D/2)Y_1(k_2 d/2), \\
G_2 &= J_0(k_2 d/2)Y_0(k_2 D/2) - J_0(k_2 D/2)Y_0(k_2 d/2).
\end{aligned} \quad (12)$$

Note that in the case of the absence of the surface crystalline layer, $t_c = 0$ and $d = D$, it follows from equations (13) that $F_1 = F_4 = G_2 = 0$ and $F_2 = F_3$, and equation (12) coincides with the usual expression for the impedance of amorphous wire at high-frequency approximation [10,11].

## 3. Results and discussion

Figure 1 shows the magnetization curves in the amorphous region calculated by means of equation (4) for different values of the bias field $H_b$. At low $H_b$, the magnetization curves show the asymmetric hysteretic behaviour. It follows from figure 1 that there are two solutions of equation (4) with different equilibrium magnetization angles at low external fields. Within this field range, the so-called bamboo domain structure may appear in the amorphous region. This domain structure consists of circular domains with the opposite magnetization direction [13,14]. The magnetization curves shift to the positive external field with the increase in the bias field. Note that the magnetization curves presented in figure 1 are similar to those calculated for the case of the asymmetric GMI due to the dc bias current [10,15], since in the studied case the configuration of the bias field is the same. If the bias field $H_b$ exceeds the threshold value of $H_a\sin\psi$, the hysteretic magnetization curve transfers to the asymmetric non-hysteretic one [15] (see figure 1).

The effect of the bias field $H_b$ on the field dependence of the impedance $Z$ is shown in figure 2. To calculate the impedance, we assume that the bamboo domain structure exists at low external fields and average the impedance response over the domain structure. The averaging procedure is similar to that described in [7]. It follows from figure 2 that the field dependence of the impedance shows the asymmetric two-peak behaviour. At low $H_b$, the asymmetry increases with the bias field, the peak at the negative field decreases and the peak at the positive field increases. When the bias field $H_b$ exceeds $H_a \sin\psi$, the domain structure disappears and the peak at the positive field increases sharply. With further growth of the bias field, the asymmetry between peaks becomes smaller (see figure 2).

Shown in figure 3 is the field dependence of the impedance at different values of the anisotropy axis angle $\psi$. The impedance response and asymmetry between the peaks increase with the decrease in the anisotropy axis angle. However, the asymmetry disappears in the case of the circular anisotropy, $\psi = 0$, when the field dependence of $Z$ becomes symmetric (see figure 3). In this case, the directions of the uniaxial anisotropy field and the bias field coincide, and the bias field does not lead to asymmetry in the dc magnetization. This result coincides with that obtained in the study of the asymmetric GMI due to the dc bias current in amorphous wires with circular anisotropy [10]. Hence, the existence of the helical magnetic anisotropy in the amorphous region is required to observe the asymmetry in the field dependence of the impedance.

Figure 4 illustrates the effect of the crystalline layer thickness $t_c$ on the frequency dependence of the wire impedance. The increase in the impedance with the frequency is related to the decrease in the skin depth in the amorphous region. The impedance field dependence remains the same for all values of $t_c$, since the permeability of the crystalline layer is independent of the external field. It follows from figure 4 that the magnitude of the impedance depends significantly on the thickness of the crystalline layer, and this dependence is more pronounced at high frequencies. This is due to the increase in the contribution from the crystalline layer to the impedance with the growth of the frequency. As a result, for the wire with a thick crystalline layer, the increase in the impedance at high frequencies is small, and the asymmetry between the peaks decreases (see figure 4).

The frequency dependence of the difference between two peaks of the impedance, $\Delta Z_{pp}$, is presented in figure 5. The asymmetry factor $\Delta Z_{pp}$ increases with the frequency, achieves a maximum and then decreases. Such frequency dependence of the asymmetry factor is related to the influence of the surface crystalline layer on the impedance. It follows from figure 5 that the frequency, at which the asymmetry factor has a maximum, increases with the decrease in the crystalline layer thickness. Note that the similar behaviour of the frequency dependence of the asymmetry factor $\Delta Z_{pp}$ with the maximum at 4 MHz has been observed in experiment [9].

Note that the impedance field dependence observed in experiment [9] shows the asymmetric two-peak behaviour at all frequencies within the range 0.5 to 10 MHz. This result differs from that obtained in the study of the exchange-biased asymmetric GMI in field-annealed Co-based amorphous ribbons. In the amorphous ribbons, the GMI response changes drastically with the frequency. At low frequencies, the single-peak 'GMI valve' has been observed, whereas at high frequencies the GMI profile exhibits asymmetric two-peak behaviour [5]. The transition from the single-peak to two-peak field dependence of the impedance can be explained by the effect of the domain-walls motion on the permeability [7]. At low frequencies, the main contribution to the permeability is due to the domain-walls motion, which results in the single-peak behaviour. At high frequencies, the domain-walls motion is damped by eddy currents, the magnetization rotation process determines the permeability and the impedance field dependence has the two-peak behaviour [7]. Most probably, in Joule-heated wires, the single-peak behaviour of the impedance has not been observed in [9] due to low relaxation frequency for the domain-walls motion. Indeed, the relaxation frequency is inversely proportional to the static domain-wall susceptibility, the wire diameter and the domain size [14]. Simple estimations show that for the domain width of 10 μm and the static susceptibility of the order of $10^3$, the relaxation frequency is less than 100 kHz. Thus, the contribution from the domain-walls motion to the permeability is insignificant to describe the exchange-biased asymmetric GMI in thick Joule-heated wires.

In the model proposed, we assume that the helical uniaxial anisotropy exists in the whole amorphous region. In real amorphous wires, the distribution of easy axes is determined by the residual stresses appearing during the wire fabrication. This results in the peculiar core−shell domain structure in Co-based amorphous wires, which consists of the longitudinally magnetized core and the shell with the circular or helical anisotropy [16]. Although the shell gives the main contribution to the GMI at sufficiently high frequencies, it has been demonstrated that the effect of the longitudinally magnetized core on the impedance may be essential at low external fields [17,18]. Note that corresponding modifications taking into account the core can be made in the framework of the present approach by considering an additional region with the longitudinal anisotropy near the wire axis.

It should be noted also that it is assumed above that the effective bias field $H_b$ has the opposite direction with respect to the unidirectional anisotropy field $H_u$ in the crystalline layer, that is, we consider the case of the antiferromagnetic coupling. The antiferromagnetic coupling conception allows one to explain the main features of the field and frequency dependences of the exchange-biased asymmetric GMI in field-annealed Co-based amorphous ribbons [7,8,19]. Nevertheless, some experimental data denotes the presence of the ferromagnetic coupling in stress-field annealed Co-based amorphous ribbons (the bias field $H_b$ has the same direction as the unidirectional anisotropy field $H_u$) [20]. This contradiction does not exist under experimental conditions for Joule-heated Co-based wires [9]. In order to investigate the bias field strength, the dc bias current has been applied to the wire [9]. It has been observed that applying low positive dc bias current (the direction of the bias current coincides with that of the annealing current) results in the decrease in the asymmetry in the GMI response. At some value of the bias current the asymmetry disappears, and at higher bias currents, the peak at the negative field becomes higher than that at the positive field. This result demonstrates that the exchange bias field has the opposite direction with respect to the unidirectional anisotropy field induced by the annealing current, and the coupling between two phases is the antiferromagnetic one.

## 4. Conclusions

A model for the exchange-biased asymmetric GMI in Joule-heated amorphous wires is developed. It is assumed that Joule heating induces the hard magnetic surface crystalline layer with the circular unidirectional anisotropy. In the framework of the single-domain approximation, the field distribution within the wire and the impedance are found. It is demonstrated that the antiferromagnetic coupling between the crystalline layer and the amorphous region results in the asymmetric GMI response. The calculated field and frequency dependences of the impedance are in a good qualitative agreement with experimental results [9].

**Acknowledgments**

This work was supported by the Korea Science and Engineering Foundation through ReCAMM. N.A. Buznikov wishes to thank the support of the Brain Pool Program.

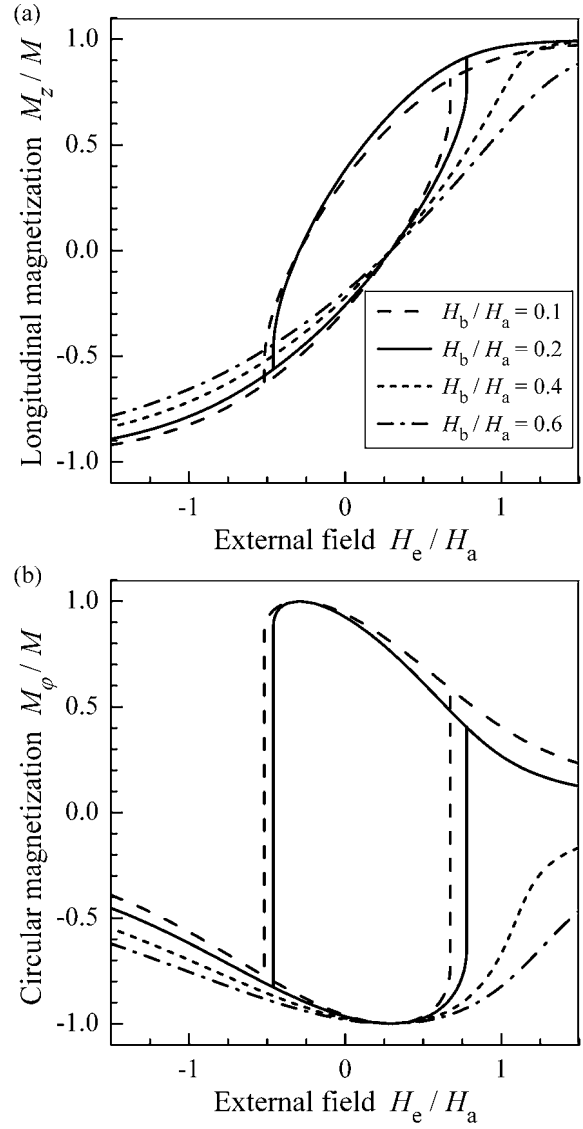

**Figure 1.** The dependences of (*a*) longitudinal, $M_z = M\sin\theta$, and (*b*) circular, $M_\varphi = M\cos\theta$, magnetization components on the external field at $\psi = 0.1\pi$ and different values of the bias field $H_b$.

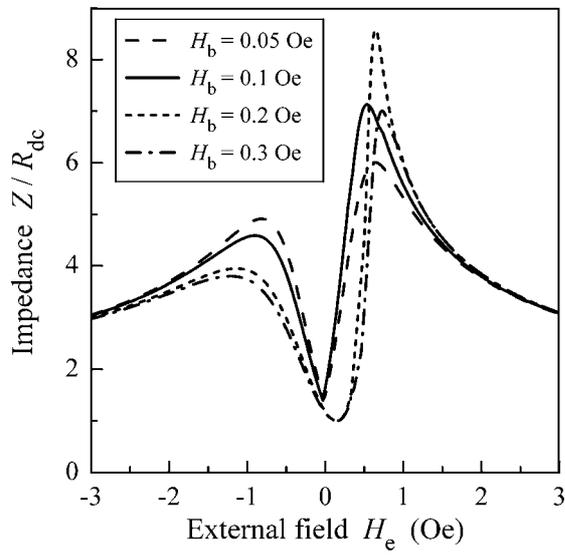

**Figure 2.** The dependence of the impedance $Z$ on the external field $H_e$ at $f = \omega/2\pi = 500$ kHz and different values of the bias field $H_b$. Parameters used for calculations are $D = 115$ μm, $t_c = 1$ μm, $M = 600$ G, $H_a = 0.5$ Oe, $\psi = 0.1\pi$, $H_u = 300$ Oe, $\sigma = 10^{16}$ s$^{-1}$, $\kappa = 0.1$.

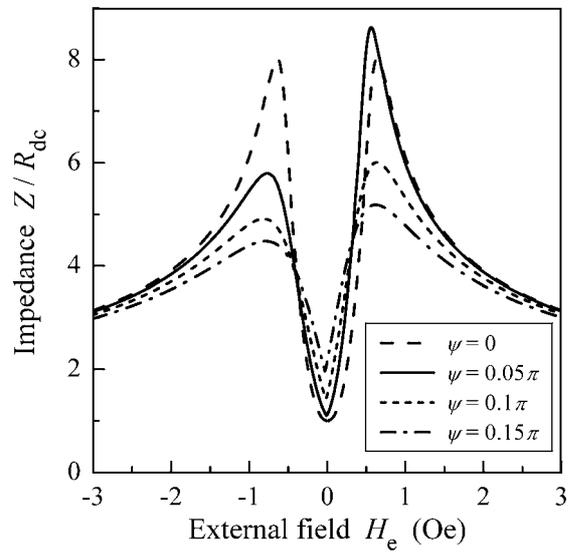

**Figure 3.** The dependence of the impedance $Z$ on the external field $H_e$ at $f = 500$ kHz and different values of the anisotropy axis angle $\psi$. Parameters used for calculations are $D = 115$ μm, $t_c = 1$ μm, $M = 600$ G, $H_a = 0.5$ Oe, $H_b = 0.05$ Oe, $H_u = 300$ Oe, $\sigma = 10^{16}$ s$^{-1}$, $\kappa = 0.1$.

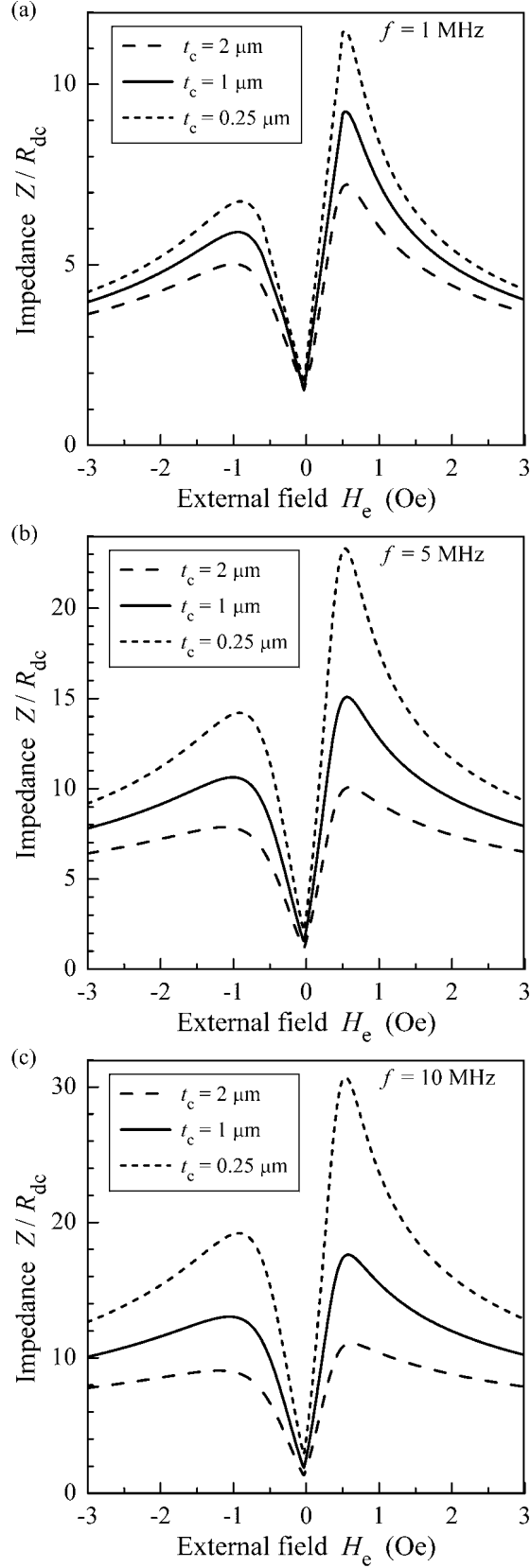

**Figure 4.** The dependence of the impedance $Z$ on the external field $H_e$ at different frequencies and different thickness of the crystalline layer $t_c$: (*a*) $f = 1$ MHz; (*b*) $f = 5$ MHz; (*c*) $f = 10$ MHz. Parameters used for calculations are $D = 115$ μm, $M = 600$ G, $H_a = 0.5$ Oe, $H_b = 0.1$ Oe, $\psi = 0.1\pi$, $H_u = 300$ Oe, $\sigma = 10^{16}$ s$^{-1}$, $\kappa = 0.1$.

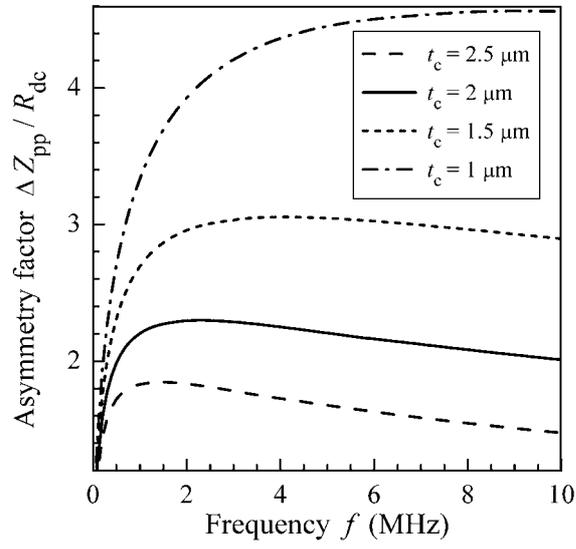

**Figure 5.** The dependence of the asymmetry factor $\Delta Z_{pp}$ on the frequency $f$ at different thickness of the crystalline layer $t_c$. Parameters used for calculations are $D = 115$ μm, $M = 600$ G, $H_a = 0.5$ Oe, $H_b = 0.1$ Oe, $\psi = 0.1\pi$, $H_u = 300$ Oe, $\sigma = 10^{16}$ s$^{-1}$, $\kappa = 0.1$.